\documentclass[aps,prl,reprint,superscriptaddress,floatfix]{revtex4-1}

\usepackage{amsmath,amsfonts,amssymb}
\usepackage{mathrsfs}
\usepackage{xparse}
\usepackage{physics}
\usepackage{booktabs}
\usepackage{graphicx}
\usepackage[utf8]{inputenc}
\usepackage[english]{babel}
\usepackage{hyperref}

\graphicspath{{./figures/}}
\renewcommand{\mathbf}{\boldsymbol}
\DeclareDocumentCommand\term{mmg}{%
  {$^{#1}$#2%
  \IfNoValueF {#3} {$_{#3}$}%
  }%
}

\makeatletter
\newcommand\etal{\emph{et al}\@ifnextchar.{}{.\ }}
\newcommand\eryso{Er$^{3+}$:Y$_2$SiO$_5$\@ifnextchar.{}{\@ifnextchar,{}{ }}}
\newcommand\yso{Y$_2$SiO$_5$\@ifnextchar.{}{\@ifnextchar,{}{ }}}
\newcommand\peryso{$^{167}$Er$^{3+}$:Y$_2$SiO$_5$\@ifnextchar.{}{\@ifnextchar,{}{ }}}
\makeatother

\newcommand{\jevsays}[1]{\textbf{Jevon says: #1}}
\newcommand{\stesays}[1]{\textbf{Stephen says: #1}}
\newcommand{\sebsays}[1]{\textbf{Sebastian says: #1}}
\newcommand{\jelenasays}[1]{\textbf{Jelena says: #1}}

\renewcommand{\jevsays}[1]{}
\renewcommand{\stesays}[1]{}
\renewcommand{\sebsays}[1]{}
\renewcommand{\jelenasays}[1]{}

\begin{document}
\title{Extending Phenomenological Crystal-Field Methods to $C_1$ Point-Group Symmetry: Characterization of the Optically-Excited Hyperfine Structure of $^{167}$Er$^{3+}$:Y$_2$SiO$_5$}

\author{S. P. Horvath}
\email[Corresponding author: ]{sebastian.horvath@gmail.com}
\affiliation{School of Physical and Chemical Sciences, University of Canterbury, PB 4800, Christchurch 8140, New Zealand}
\affiliation{Department of Physics, University of Otago, PB 56, Dunedin 9016, New
  Zealand}
\affiliation{The Dodd-Walls Centre for Photonic and Quantum Technologies, New Zealand}
\author{J. V. Rakonjac}
\author{Y.-H. Chen}
\author{J.~J.~Longdell}
\affiliation{Department of Physics, University of Otago, PB 56, Dunedin, New
  Zealand}
\affiliation{The Dodd-Walls Centre for Photonic and Quantum Technologies, New Zealand}
\author{P. Goldner}
\affiliation{Chimie ParisTech, PSL University, CNRS, Institut de Recherche de Chimie
Paris, 75005 Paris, France}
\author{J.-P. R. Wells}
\author{M. F. Reid}
\email[Corresponding author: ]{mike.reid@canterbury.ac.nz}
\affiliation{School of Physical and Chemical Sciences, University of Canterbury, PB 4800, Christchurch 8140, New Zealand}
\affiliation{The Dodd-Walls Centre for Photonic and Quantum Technologies, New Zealand}

\date{\today}

\begin{abstract}
  We show that crystal-field calculations for $C_1$ point-group
  symmetry are possible, and that such calculations can be performed
  with sufficient accuracy to have substantial utility for rare-earth
  based quantum information applications.  In particular, we perform
  crystal-field fitting for a $C_1$-symmetry site in
  $^{167}$Er$^{3+}$:Y$_2$SiO$_5$. The calculation simultaneously
  includes site-selective spectroscopic data up to 20,000\,cm$^{-1}$,
  rotational Zeeman data, and ground- and excited-state hyperfine
  structure determined from high-resolution Raman-heterodyne
  spectroscopy on the 1.5 $\mu$m telecom transition. We achieve an
  agreement of better than 50 MHz for assigned hyperfine
  transitions. The success of this analysis opens the possibility of
  systematically evaluating the coherence properties, as well as transition
  energies and intensities, of any rare-earth ion doped into Y$_2$SiO$_5$.

\end{abstract}

\maketitle

Over the last decade, substantial progress has been made towards realizing practical quantum information processing hardware using solid-state rare-earth ion based materials. Key areas of advancement have been optical quantum memories \cite{deRiedmatten2008,hedges2010,jobez_coherent_2015,gundogan2015,zhong_optically_2015,zhong2017,rancic2018,lauritzen2010,laplane2017,seri2017}, quantum-gate implementations \cite{longdell_experimental_2004,rippe2008}, single photon sources \cite{kolesov2012,utikal2014,thompson2018}, and microwave-to-optical photon modulators \cite{xavi2015}. To date, one of the host materials of choice for these applications has been yttrium orthosilicate (Y$_2$SiO$_5$). The reason for this is twofold: first, yttrium has a very small nuclear magnetic moment, while isotopes of Si and O with nonzero nuclear spin  have very low natural abundances. At cryogenic temperatures, nuclear spin flips are the primary source of decoherence in rare-earth ion doped materials, resulting in \yso based systems having outstanding coherence properties. The second reason is that the rare-earth substitutional site in \yso has a $C_1$ point-group symmetry; this leads to highly admixed wave functions enabling efficient and diverse optical pumping schemes \cite{rippe2005,lauritzen2008,longdell_experimental_2004}. 

The formulation of accurate models for the hyperfine structure of $C_1$ point-group symmetry sites is highly nontrivial; however, they are an invaluable tool for a number of practical applications. For example, the availability of the spin Hamiltonian for $^{151}$Eu$^{3+}$:\yso allowed for a computational search for magnetic field orientations exhibiting a near-zero gradient with respect to hyperfine energy levels. This is the basis of the Zero-First-Order-Zeeman (ZEFOZ) technique, which was essential to the experimental demonstration of a coherence time of six hours \cite{zhong_optically_2015}. However, spin Hamiltonian models are restricted to specific electronic levels of a single ion-host combination. This results in considerable practical challenges, especially for the structure of the excited-state electronic levels, which can often only be probed using experiments that conflate excited and ground state splittings. In this work, we avoid the shortcomings of spin Hamiltonians by developing a method to fit a crystal-field Hamiltonian for erbium doped \yso. 

Crystal-field methods have been essential to the development of rare-earth optical applications, such as phosphors and lasers \cite{krupke1971,judd1988,Gorller-Walrand1998}. However, the lack of symmetry (i.e. $C_1$ symmetry) of rare-earth substitutional sites in \yso hindered the application of crystal-field modeling to this material, despite its ubiquity as a host in quantum-information applications. Previous work on $C_1$ symmetry sites was based on \emph{ab initio} calculations \cite{doualan_energy_1995,wen2014}, or used a higher-symmetry approximation to reduce the number of parameters \cite{guillot-noel_calculation_2010,sukhanov2018}. These approaches are not accurate enough to model the complex magnetic and hyperfine structure that we consider in this work. The key advantage of a crystal-field model over the spin Hamiltonian approach is that it is not restricted to a specific electronic level, but predicts the magnetic and hyperfine structure of the complete $4f$ configuration. This greatly increases the predictive power, aids the analysis of excited states, and also enables fitting to a much wider range of experimental data. Further, crystal-field modeling enables the rigorous calculation of radiative transition rates \cite{Gorller-Walrand1998,reid_electronic_2006}. 

The predictive power of crystal-field models extends considerably beyond an individual rare-earth ion. For a fixed host crystal, there exist well established parameter trends across the rare-earth series \cite{carnall_systematic_1989}. Consequently, parameters describing a specific rare-earth dopant can be extrapolated to previously unstudied rare-earth dopants in the same host. Moreover, the availability of complete $4f^{11}$ wave functions enables several novel applications, such as studying ZEFOZ points in a large magnetic field, a regime in which the spin Hamiltonian approach breaks down. A recent demonstration of coherence times exceeding one second in \peryso using a 7 T magnetic field makes this particularly relevant \cite{rancic2018}, since if such an approach were to be combined with the ZEFOZ technique, an accurate model at large field would be imperative.

In this letter we report a phenomenological crystal-field fit for one of the $C_1$ symmetry sites of \peryso. Physical properties, such as a transitions in the 1.5 $\mu$m telecommunications band and an optical homogeneous linewidth of 50 Hz \cite{sun2002} make this one of the most promising materials for rare-earth based quantum information applications. Despite extensive past characterization \cite{doualan_energy_1995,bottger2006,bottger2008,baldit_identification_2010,lauritzen_state_2008,budoyo2018,sun_magnetic_2008,chen2018}, an accurate model of the excited state hyperfine structure remains an outstanding problem. This material is, therefore, an important test case for crystal-field fitting to substitutional sites without symmetry. To achieve a unique fit, both site-selective optical as well as Zeeman and hyperfine data were required \cite{horvath_caf2_2018}, which were available from the literature \cite{doualan_energy_1995,sun_magnetic_2008,chen2018}. This was complemented with targeted Raman-heterodyne measurements to obtain high-precision hyperfine splittings of the ground and \term{4}{I}{13/2}$Y_1$ excited states.

Y$_2$SiO$_5$ is a monoclinic crystal with $C^6_{2h}$ space group symmetry. The yttrium ions occupy two crystallographically distinct sites, each with $C_1$ point-group symmetry, referred to as site 1 and site 2 \cite{maksimov1971crystal}. Because the wavelength tuning range of our laser, this work is focused on site 1. Y$_2$SiO$_5$ has three perpendicular optical-extinction axes: the crystallographic $b$ axis, and two mutually perpendicular axes labeled $D_1$ and $D_2$. We follow the convention of identifying these axes as $z$, $x$, and $y$, respectively \cite{sun_magnetic_2008}. 

The complete Hamiltonian appropriate for modeling the $4f^n$ configuration reads
\begin{equation}
  H = H_{\mathrm{FI}} + H_{\mathrm{CF}} + H_{\mathrm{Z}} + H_{\mathrm{HF}} + H_{\mathrm{Q}}.
  \label{eqn:h_defn}
\end{equation} 
The terms in the above equation represent the following interactions: the free-ion contribution, the crystal-field interaction, the Zeeman term, the nuclear magnetic dipole hyperfine interaction, and the nuclear quadrupole interaction. We use the usual free-ion Hamiltonian with the following parameters: $E_0$ accounting for a constant configurational shift, $F^k$, the Slater parameters characterizing aspherical electrostatic repulsion, and $\zeta$, the spin-orbit coupling constant. Furthermore, we also include terms that parametrize two- and three-body interactions as well as higher-order spin-dependent effects; for a more detailed description, the reader is referred to the review by Liu \cite{liu_electronic_2006}. The most general crystal-field Hamiltonian has the form
\begin{equation}
  H_{\text{CF}} = \sum_{k,q} B^k_q C^{(k)}_q,
  \label{eqn:hcf_c1}
\end{equation}
for $k = 2, 4, 6$ and $q = -k \cdots k$. The $B^k_q$ parameters are the crystal-field expansion coefficients and C$^{(k)}_q$ are spherical tensor operators using Wybourne's normalization \cite{wybourne_spectroscopic_1965}.
We write nonaxial ($q \ne 0$) $B^k_q$ parameters as complex numbers. In this convention the $\pm q$ parameters are related by $(B^k_{q})^* = (-1)^q B^k_{-q}$ \cite{NN89a,newman2007crystal}. For the remaining terms in Eq.\ \eqref{eqn:h_defn} we note that $H_{\mathrm{HF}}$ and $H_{\mathrm{Q}}$, respectively, contain coupling constants $A$ and $Q$ that must be determined from experiment, while $H_{\mathrm{Z}}$ has no free parameters. For a detailed description of these terms, and the evaluation of their matrix elements, the reader is referred to Refs.\ \cite{mcleod_intensities_1997,guillot-noel_calculation_2010}.  

High precision magnetic and hyperfine interactions are generally expressed using the spin Hamiltonian formalism \cite{macfarlane_coherent_1987}.  For a Kramers ion with nonzero nuclear spin, this Hamiltonian has the form \cite{abragam_electron_1970}
\begin{equation}
  \mathscr{H} = \beta_e \mathbf{B} \cdot \mathbf{g} \cdot \mathbf{S} + \mathbf{I} \cdot \mathbf{A} \cdot \mathbf{S} + \mathbf{I} \cdot \mathbf{Q} \cdot \mathbf{I} - \beta_n g_n \mathbf{B} \cdot \mathbf{I},
  \label{eqn:sh}
\end{equation}
where $\beta_e$ is the Bohr magneton, $\mathbf{B}$ is an external field vector, $\mathbf{g}$ is the $g$ tensor, $\mathbf{A}$ is the hyperfine tensor, and $\mathbf{Q}$ is the electric-quadrupole tensor. Further, $\mathbf{S}$ and $\mathbf{I}$ are vectors of electronic and nuclear spin operators, respectively. $\beta_n$ and $g_n$ are the nuclear magneton and nuclear $g$ factors, respectively. For the magnetic field values considered here, the nuclear Zeeman interaction is less than 2 MHz, and since the uncertainty of $\mathbf{A}$ and $\mathbf{Q}$ is $O(20 \text{ MHz})$ \cite{chen2018} this interaction is neglected.

For the initial phase of our fitting, we use a projection from the crystal-field Hamiltonian to the spin Hamiltonian, so that we can fit to spin-Hamiltonian parameters. This projection has the form
\begin{equation}
  A_\mathrm{SH} = V^\dag A V,
  \label{eqn:sh_proj}
\end{equation}
for operator $A$ and spin Hamiltonian effective operator $A_{\mathrm{SH}}$. Here, $V$ are the eigenvectors one obtains by diagonalizing $H_{\mathrm{FI}} + H_{\mathrm{CF}}$, which can be interpreted as the zero order contribution to the spin Hamiltonian. 

For $C_1$ symmetry this projection has some subtleties; specifically, there is a phase freedom in the matrix elements of $\mathbf{S}$ and $\mathbf{I}$ in Eq.\ \eqref{eqn:sh}. This phase freedom does not affect the eigenvalue spectrum of the spin Hamiltonian; nevertheless, a specific orientation is required in order for the parameter tensors to be symmetric. When one determines spin Hamiltonian parameter matrices from experimental data, this issue is avoided, for by choosing symmetric parameter matrices during the fitting, one implicitly fixes the phase to an appropriate value. However, when one performs the projection \eqref{eqn:sh_proj}, the value of this phase does not necessarily correspond to a symmetric parameter tensor. We mitigate this by employing a singular-value decomposition to transform the spin Hamiltonian tensors to a basis in which they are always symmetric.
For our calculations, this phase was identically zero for matrix elements of $\mathbf{I}$, and therefore, we only discuss matrix elements of $\mathbf{S}$. We consider, as an example, the Zeeman interaction term. Given the unitary matrices $\mathbf{U}$ and $\mathbf{V}$ and the diagonal matrix $\mathbf{\Sigma}$, the singular value decomposition of $\mathbf{g}$ takes the form
\begin{equation}
  \mathbf{g} = \mathbf{U} \mathbf{\Sigma} \mathbf{V}^\dag.
  \label{eqn:g_svd}
\end{equation}
Consequently, $\mathbf{U}^\dag \mathbf{g} \mathbf{V}$ is diagonal, and performing a similarity transformation with the unitary matrix $\mathbf{U}$ we obtain the symmetric tensor
\begin{equation}
  \mathbf{U} \mathbf{U}^\dag \mathbf{g} \mathbf{V} \mathbf{U}^\dag = \mathbf{g} \mathbf{V} \mathbf{U}^\dag. 
  \label{eqn:g_prime}
\end{equation}
Thus, we can define a transformed set of electronic spin operators $\mathbf{S}' = \mathbf{R} \mathbf{S}$ with $\mathbf{R} = \mathbf{V} \mathbf{U}^\dag$, leading to an SU(2) transformed spin Hamiltonian term of the form $\mathbf{B} \cdot \mathbf{g}' \cdot \mathbf{S}'$, with $\mathbf{g}' = \mathbf{g} \mathbf{V} \mathbf{U}^\dag$ symmetric. An analogous procedure can be applied to the nuclear dipole interaction term $\mathbf{I} \cdot \mathbf{A} \cdot \mathbf{S}$. 

The crystal-field fit was performed in two phases: an initial coarse fitting which excluded high-resolution Raman-heterodyne data, and a second polishing phase where hyperfine transition data were iteratively added. The initial fitting employed site-selective excitation and fluorescence data from Doualan \etal \cite{doualan_energy_1995}, while simultaneously including the $g$ tensor of the \term{4}{I}{13/2}$Y_1$ level reported by Sun \etal \cite{sun_magnetic_2008}, as well as the complete ground-state spin Hamiltonian reported by Chen \etal \cite{chen2018}. The site-selective data was, as usual in crystal-field calculations, directly fit to the eigenvalues of Eq. \eqref{eqn:h_defn}. In order to simultaneously fit to spin Hamiltonian data, the projection \eqref{eqn:sh_proj} was utilized to obtain a theoretical set of parameter matrices which could be fitted to their experimental counterparts. 

This procedure yielded a set of parameters of sufficient accuracy to identify several \term{4}{I}{13/2} hyperfine transitions in our Raman-heterodyne data and, thus, complete the coarse step of the fitting. In order to perform the polishing stage, the projection \eqref{eqn:sh_proj} was abandoned, and instead, the Hamiltonian \eqref{eqn:h_defn} was evaluated for a range of magnetic field values to directly obtain eigenvalues describing both the hyperfine structure as well as the site-selective data. This has the advantage that Raman-heterodyne data could be added step-by-step as transitions were identified. In order to ease the computational burden the calculations of hyperfine states were performed using a truncated basis using the intermediate-coupling method described by Carnall \etal \cite{carnall_systematic_1989}. All software used to perform these calculations is available from \cite{horvath_pycf}. 

Raman-heterodyne spectroscopy was performed for two separate frequency regions. Between 0-100 MHz we used an RF coil, and between 600-1200 MHz a tunable aluminium single-loop single-gap resonator was used. Samples were cooled using a home built cryostat (containing a Cryomech PT405 pulsetube cooler) with an HTS-100 Ltd superconducting vector magnet to provide an arbitrarily-oriented magnetic field. The light source was a Koheras AdjustiK E15 fiber laser, operating at 1536.48 nm on resonance with the \term{4}{I}{15/2} $\to$ \term{4}{I}{13/2} transition of site 1. The sample was an isotopically purified \peryso crystal (Scientific Materials Inc.) with $^{167}$Er$^{3+}$ substituted for Y$^{3+}$ ions at a 50 ppm level. For a more detailed description of Raman-heterodyne spectroscopy, as well as the experimental setup and methods, the reader is referred to Ref.\ \cite{rakonjac2018}, which employed the same Raman-heterodyne setup to identify transitions with long spin-coherence times.

\begin{figure}[tb!]
\includegraphics[width=\linewidth]{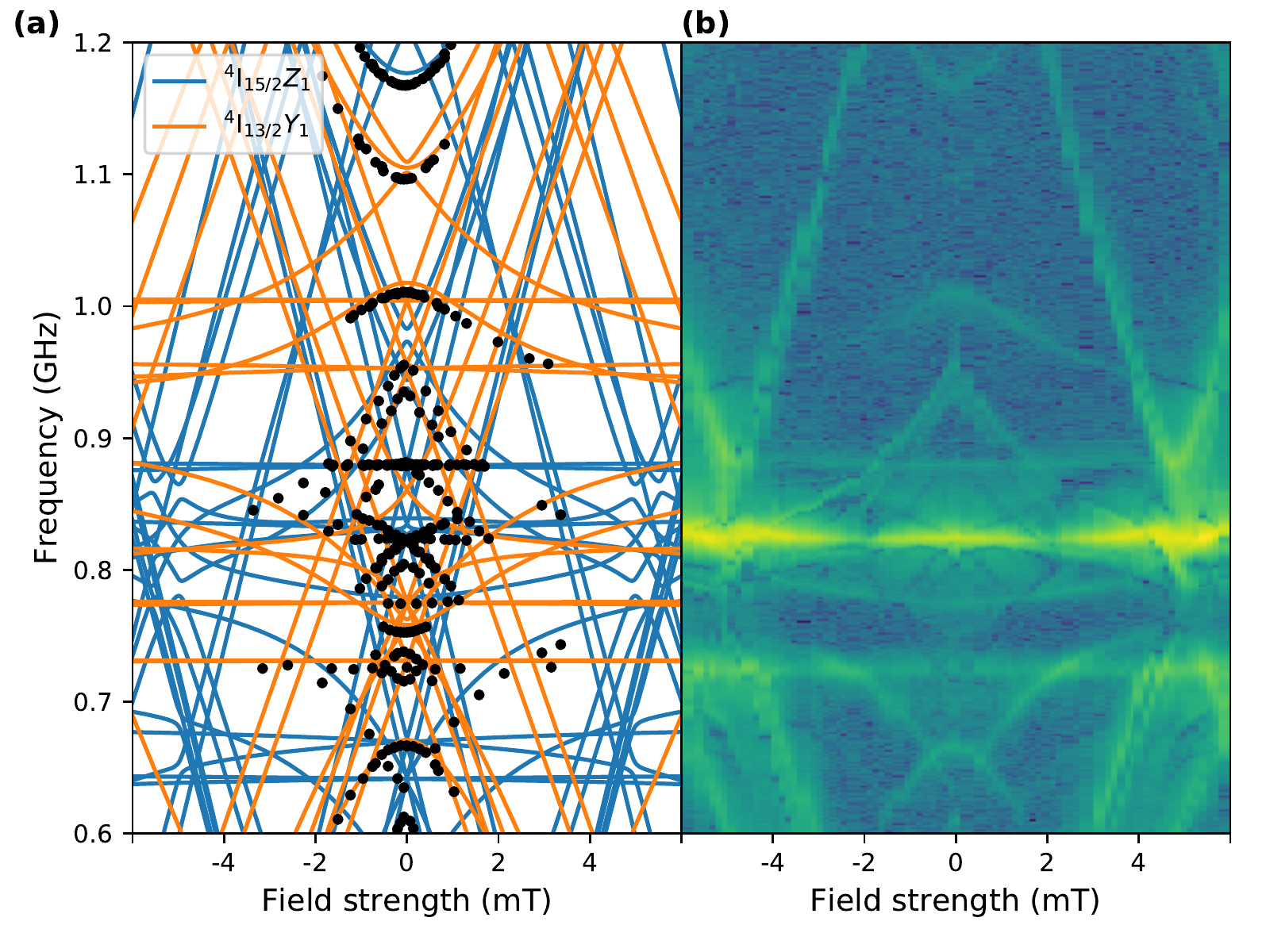}
\caption{\label{fig:cfvs41} (a) Raman-heterodyne data (black markers) showing hyperfine transitions of the first levels of both the $^4$I$_{15/2}$ and $^4$I$_{13/2}$ multiplets of site 1 in Er$^{3+}$:Y$_2$SiO$_5$, superimposed with predictions from our crystal-field model. The magnetic field was varied in the direction of the $D_2$ axis. (b) A Raman-heterodyne scan of this region; the color map uses a linear scale of arbitrary intensity with yellow/green indicating a resonance condition.}% for a microwave transition in either the ground or the excited state.}
\end{figure}
Figure \ref{fig:cfvs41} shows the hyperfine transitions of both the ground and excited states with resonances in the 600-1200 MHz region with respect to a small change in magnetic field along the $D_2$ axis. These measurements are for site 1 of \peryso. Most transitions were studied in further detail using higher resolution scans over restricted subfrequencies to provide detailed curvatures for comparison with our model. Furthermore, low frequency data at 85 MHz included curvatures with respect to an external field along the $D_1$, $D_2$, and $b$ axes. The maximum deviation of any Raman-heterodyne transition that we directly fit to was 15 MHz. For a few transitions the assignments remained ambiguous due to the closely spaced spectral lines. The maximum difference between an observed transition and its theoretical prediction was $50$ MHz, approximately 1\% of the span of the hyperfine levels. We note that, using our final transition assignments, the coarse fitting predicted \term{4}{I}{13/2}$Y_1$ transition frequencies to within $\sim 200$ MHz of their measured values. Thus, while using only ground-state hyperfine data enabled the prediction of the excited-state hyperfine transitions with reasonable accuracy, further fitting was required to obtain an optimal model. 

Given that similar ground-state data is available for crystallographic site 2 of \peryso \cite{chen2018}, a model with comparable accuracy to the coarse fitting presented here should, in principle, be possible; however, such an analysis is beyond the scope of this work.
\begin{figure}[tb!]
\includegraphics[width=\linewidth]{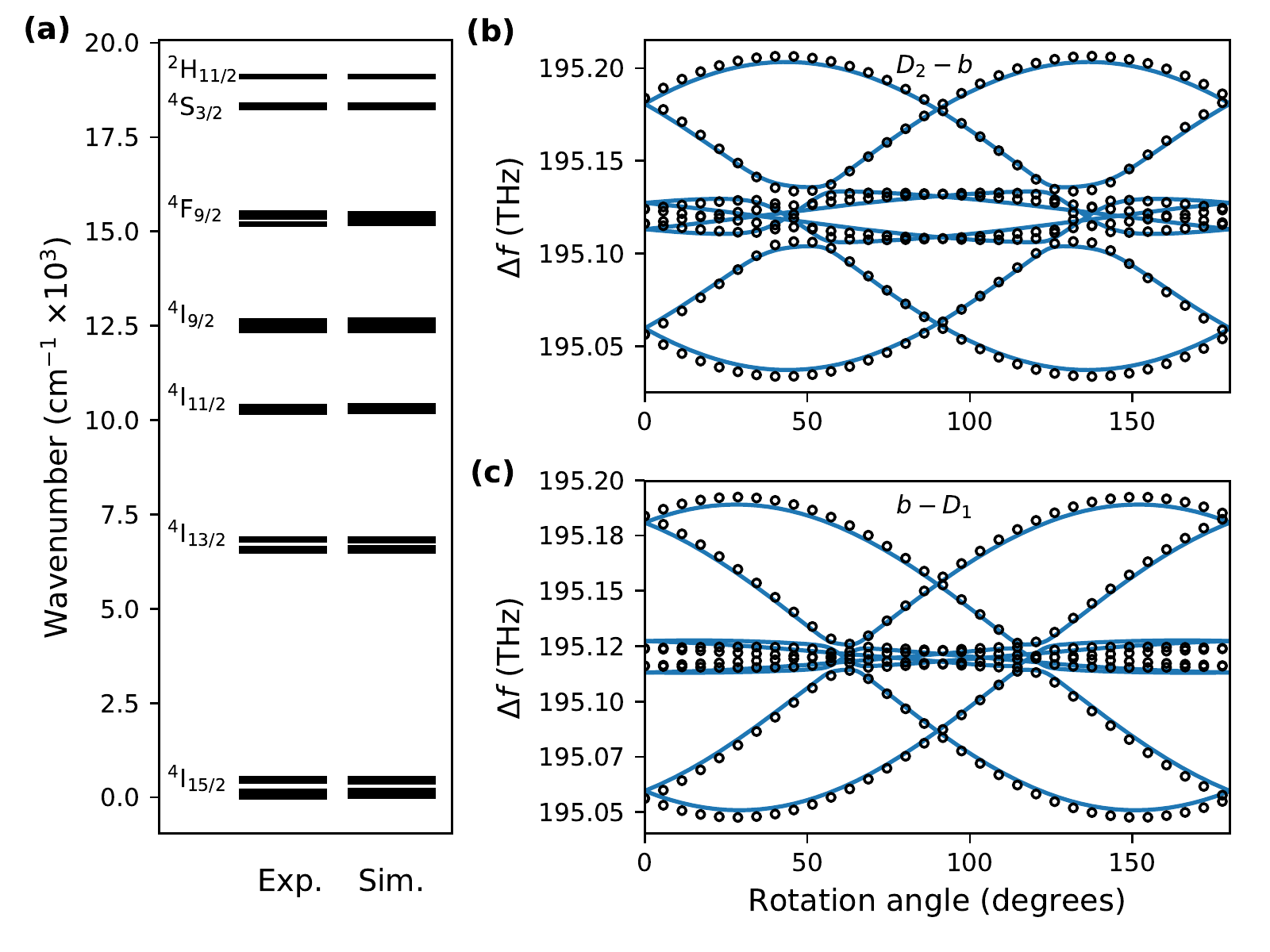}
\caption{\label{fig:stark_mag_comb} (a) Experimental and simulated crystal-field level splittings up to $^2$H$_{11/2}$ for site 1 of Er$^{3+}$:Y$_2$SiO$_5$; experimental values are from Doualan \emph{et al.} \cite{doualan_energy_1995}. (b) and (c) Rotation patterns for optical transitions between \term{4}{I}{15/2}$Z_1$ and \term{4}{I}{13/2}$Y_1$. Circles denote the predictions using $g$-tensor data from Sun \emph{et al.} \cite{sun_magnetic_2008}, while the solid lines correspond to our crystal-field model. The magnetic field magnitude used was $0.484$ T, and the labels $D_2 - b$ and $b - D_1$ indicate the rotation planes using the standard orthogonal axes notation for Y$_2$SiO$_5$.}
\end{figure}

In Fig.\ \ref{fig:stark_mag_comb} we show the spread of electronic data for energies up to 20,000\,cm$^{-1}$ and include detailed magnetic rotation data of the transition between the lowest \term{4}{I}{13/2} level and the ground state, with splittings on the order of 200 GHz \cite{sun_magnetic_2008}. In Table \ref{tab:cf_params} we present the complete set of parameters determined from our calculation. We note that since the Zeeman and hyperfine splittings are much smaller in magnitude and are determined with higher accuracy than the crystal-field level energies, they were given more weight in the parameter fit.

The parameter uncertainties shown in Tab.\ \ref{tab:cf_params} were estimated using the Markov Chain Monte-Carlo technique. In the Supplemental Material, which includes Refs.\ \cite{wales_global_1997,rowan1990,nlopt,aster2011}, we provide a more detailed description of the fitting procedure. We also include all predicted crystal-field level energies up to \term{2}{H}{11/2} for a direct comparison with experimental values from Ref.\ \cite{doualan_energy_1995}. Furthermore, the Raman-heterodyne data for the $0-120$ MHz frequency window is presented and plotted together with the corresponding theoretical transition energies.
\begin{table}[tb!]
  \centering
  \caption{\label{tab:cf_params}%
  Fitted values for the free-ion and crystal-field parameters for site 1 of Er$^{3+}$:Y$_2$SiO$_5$. The Judd and Tree's parameters, which are not included here, were fixed to the values obtained for Er$^{3+}$:LaF$_3$ by Carnall \emph{et al}.\ \cite{carnall_systematic_1989}. 
}
  \begin{tabular}{ccc}
    \hline \hline
    Parameter & Fitted value (cm$^{-1}$) & Uncertainty (cm$^{-1}$) \\
    \hline
    $E_0$ & 35503.5 & 19.8 \\
    $\zeta$ & 2362.9 & 1.8 \\
    $F^2$ & 96029.6 & 183.7 \\
    $F^4$ & 67670.6 & 223.2 \\
    $F^6$ & 53167.1 & 263.7 \\
    $B^2_0$ & -149.8 & 5.4 \\
    $B^2_1$ & 420.6+396.0i & 3.1+1.3i \\
    $B^2_2$ & -228.5+27.6i & 1.8+3.4i \\
    $B^4_0$ & 1131.2 & 30.4 \\
    $B^4_1$ & 985.7+34.2i & 7.0+6.7i \\
    $B^4_2$ & 296.8+145.0i & 9.0+4.1i \\
    $B^4_3$ & -402.3-381.7i & 9.7+8.9i \\
    $B^4_4$ & -282.3+1114.3i & 13.4+12.0i \\
    $B^6_0$ & -263.2 & 3.1 \\
    $B^6_1$ & 111.9+222.9i & 1.5+3.9i \\
    $B^6_2$ & 124.7+195.9i & 2.1+3.8i \\
    $B^6_3$ & -97.9+139.7i & 5.1+9.7i \\
    $B^6_4$ & -93.7-145.0i & 4.1+3.0i \\
    $B^6_5$ & 13.9+109.5i & 2.0+6.1i \\
    $B^6_6$ & 3.0-108.6i & 8.6+2.4i \\
    $A$ & 0.005466 & 0.000003 \\
    $Q$ & 0.0716 & 0.0003 \\
    \hline \hline
  \end{tabular}
\end{table}
The Zeeman and hyperfine tensors for both the $Z_1$ and $Y_1$ electronic levels are tabulated in the Supplemental Material.  

In the crystal-field analysis, 34 parameters (five free-ion parameters, 27 crystal-field parameters, and two hyperfine parameters) are fitted to 95 data points (enumerated in the Supplemental Material). By comparison, two separate spin Hamiltonians, requiring, in total, 34 parameters, would be required for a conventional analysis of the two states. The advantage of our approach is that a fit to the ground state hyperfine data yields a prediction of the excited state hyperfine structure. This enables simultaneous fitting to both ground and excited state data to obtain a high-precision $4f^{11}$ Hamiltonian.

In conclusion, we have demonstrated a crystal-field fit for a rare-earth substitutional site with no symmetry. This enabled us to accurately characterize the hyperfine structure of the ground state and all excited state levels of \peryso, allowing modeling of optical pumping schemes via the 1.5 $\mu$m (or other) transitions, as well as high-field ZEFOZ applications. With suitable scaling, the crystal-field parameters are also applicable to other ions in Y$_2$SiO$_5$, opening the possibility of identifying promising transitions prior to extensive experimental investigation. 

The authors wish to acknowledge the use of New Zealand eScience Infrastructure (NeSI) high performance computing facilities as part of this research and financial support from the Marsden Fund of the Royal Society of New Zealand through Contract No. UOO1520. S.P.H.\ acknowledges financial support in the form of a Canterbury Scholarship by the University of Canterbury. P.G.\ would like to thank the University of Canterbury for support in the form of an Erskine Fellowship.

%

%\bibliography{eryso_crystal_field}

\end{document}

\title{Supplemental Material for: \\
``Extending Phenomenological Crystal-Field Methods to $C_1$ Point-Group
Symmetry: Characterization of the Optically Excited Hyperfine Structure of
$^{167}$Er$^{3+}$:Y$_2$SiO$_5$''
}

\author{S. P. Horvath}
\email[Corresponding author: ]{sebastian.horvath@gmail.com}
\affiliation{School of Physical and Chemical Sciences, University of Canterbury, PB 4800, Christchurch 8140, New Zealand}
\affiliation{Department of Physics, University of Otago, PB 56, Dunedin 9016, New
  Zealand}
\affiliation{The Dodd-Walls Centre for Photonic and Quantum Technologies, New Zealand}
\author{J. V. Rakonjac}
\author{Y.-H. Chen}
\author{J.~J.~Longdell}
\affiliation{Department of Physics, University of Otago, PB 56, Dunedin 9016, New
  Zealand}
\affiliation{The Dodd-Walls Centre for Photonic and Quantum Technologies, New Zealand}
\author{P. Goldner}
\affiliation{Chimie ParisTech, PSL University, CNRS, Institut de Recherche de Chimie
Paris, Paris 75005, France}
\author{J.-P. R. Wells}
\author{M. F. Reid}
\email[Corresponding author: ]{mike.reid@canterbury.ac.nz}
\affiliation{School of Physical and Chemical Sciences, University of Canterbury, PB 4800, Christchurch 8140, New Zealand}
\affiliation{The Dodd-Walls Centre for Photonic and Quantum Technologies, New Zealand}
\maketitle

\section{Spin Hamiltonian parameters} 

In Tab.\ \ref{tab:sh_params} we present the parameters for the ground state along with the \term{4}{I}{13/2}$Y_1$ excited state. These were calculated using Eq.\ (4) of the main text, with the crystal-field parameters presented in Tab.\ I of the main text. We also include parameters from literature wherever they are available. 

We note that the $\mathbf{A}$ and $\mathbf{Q}$ tensors normally obtained by fitting to experimental data include contributions from the pseudo-nuclear Zeeman and the pseudo-quadrupole interactions, respectively. In the crystal field model, these interactions are directly accounted for by the mixing of adjacent crystal-field levels. To ensure the $\mathbf{A}$ and $\mathbf{Q}$ tensors in Tab.\ \ref{tab:sh_params} match with parameters obtained from phenomenological data we perform a refining fit for $\mathbf{A}$ and $\mathbf{Q}$ using a synthetic data-set generated from our crystal-field model. The deviation between the two approaches for calculated transition frequencies was on the order of 10-20 MHz, which therefore necessitated this step. 
\begin{table}[th!]
  \caption{\label{tab:sh_params} \term{4}{I}{15/2}$Z_1$ ground and \term{4}{I}{13/2}$Y_1$ excited state spin Hamiltonian parameters determined from our crystal-field Hamiltonian for site 1 of Er$^{3+}$:Y$_2$SiO$_5$ (denoted by the prefix CF). All $\mathbf{A}$ and $\mathbf{Q}$ values are in MHz.} 
  \begin{tabular}{lcccccc}
    \hline \hline
  State & $g_{xx}$ & $g_{yy}$ & $g_{zz}$ & $g_{xy}$ & $g_{xz}$ & $g_{yz}$ \\
  \hline
  CF ground & 2.10 & 8.37 & 5.49 & -3.43 & -3.21 & 5.16 \\
  \cite{chen2018} ground & 2.90 & 8.90 & 5.12 & -2.95 & -3.56 &  5.57 \\
  CF excited & 2.04 & 4.44 & 7.94 & -2.24 & -3.40 & 5.15 \\
  \cite{sun_magnetic_2008} excited & 1.95 & 4.23 & 7.89 & -2.21 & -3.58 & 4.99 \\
   & $A_{xx}$ & $A_{yy}$ & $A_{zz}$ & $A_{xy}$ & $A_{xz}$ & $A_{yz}$ \\
  \hline
  CF ground & 200.80 & 911.27 & 586.95 & -344.23 & -362.61 & 586.95 \\
  \cite{chen2018} ground & 274.29 & 827.50 & 706.15 & -202.52 & -350.82 & 635.15 \\
  CF excited & 271.96 & 583.12 & 1058.43 & -293.37 & -447.76 & 684.97 \\
   & $Q_{xx}$ & $Q_{yy}$ & $Q_{zz}$ & $Q_{xy}$ & $Q_{xz}$ & $Q_{yz}$ \\
  \hline
  CF ground &  9.32 & -6.37 & -2.95 & 1.92 & 2.26 & -9.55 \\
  \cite{chen2018} ground & 10.40 & -5.95 & -4.44 & -9.12 & -9.96 & -14.32 \\
  CF excited & 7.47 & 0.93 & -6.50 & 3.62 & 5.54 & -9.27 \\
  \hline \hline
  \end{tabular}
\end{table}

For each crystallographic site of \yso (site 1 and site 2) there are two subclasses of ions which have different transition energies in an external field. These magnetically inequivalent sites are related by the $C^6_{2h}$ space-group symmetry. In Tab.\ \ref{tab:sh_params} we only included the $g$ tensors for a single magnetically inequivalent site; the second site can be obtained by rotating the provided parameters by
\begin{equation}
  R = \begin{bmatrix}
    -1 &  0 & 0 \\
     0 & -1 & 0 \\
     0 &  0 & 1 
   \end{bmatrix}.
\end{equation}

\section{Low frequency Raman-heterodyne data \label{sec:lf_rh}}

\begin{figure}[tbh!]
\includegraphics[width=\linewidth]{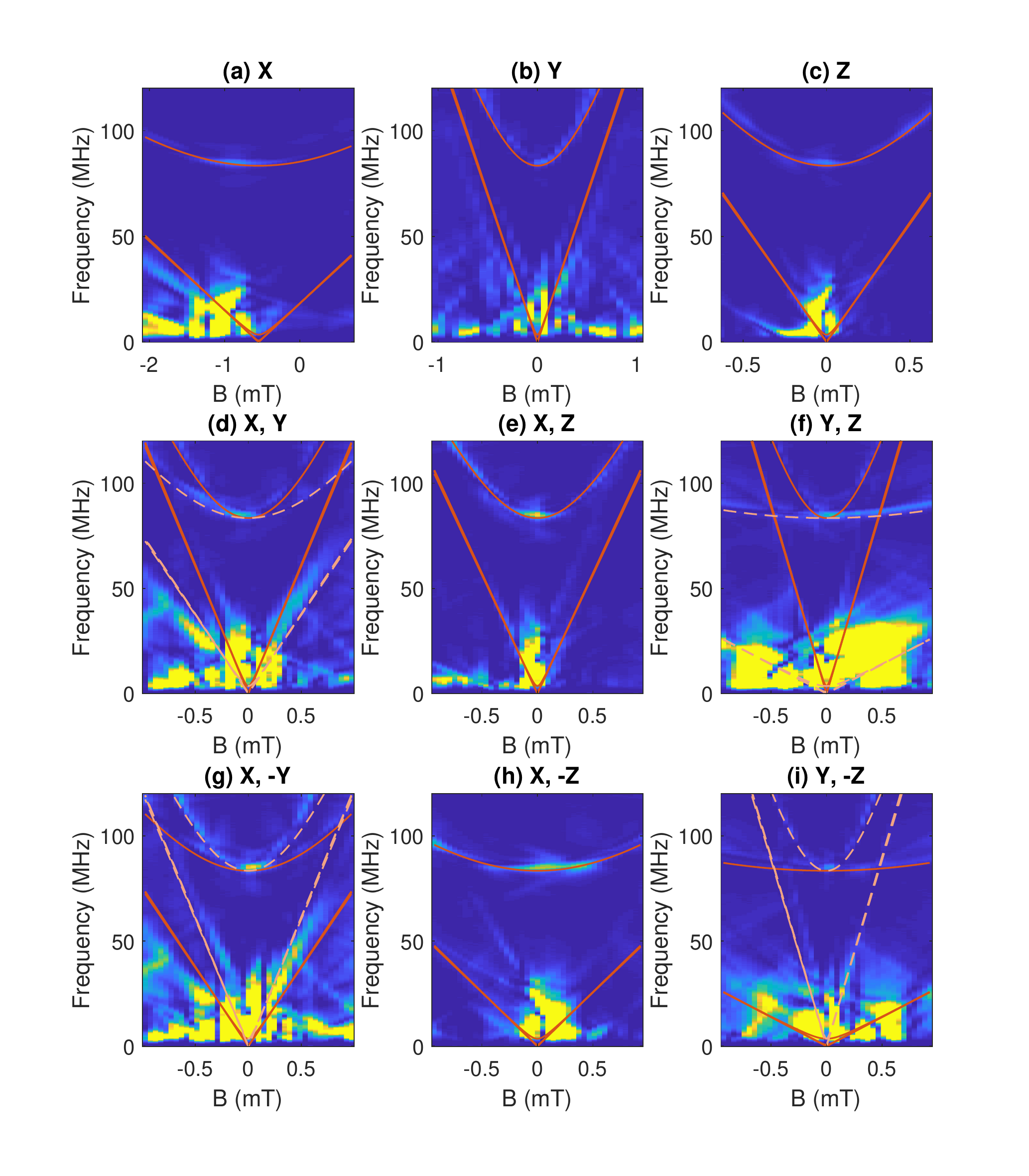}
\caption{\label{fig:85MHz} Raman-heterodyne data for nine distinct magnetic field sweeps, plotted with theoretical predictions. Field directions are indicated in subfigure titles; however, axes $X$, $Y$, and $Z$ do not map directly to the crystallographic axes. See the text for details. In the colormaps yellow/light blue indicates a resonance condition, that is, the energy of a hyperfine transition. The red lines are our prediction for the lowest energy transitions of the $^4$I$_{13/2}$$Y_1$ excited state. For directions out of the $D_1$-$D_2$ plane and not along the $b$ axis, there are two magnetically inequivalent sites. For applicable field sweeps, both magnetically inequivalent sites are plotted, and distinguished by solid and dashed lines.}
\end{figure}

\begin{figure}[tbh!]
\includegraphics[width=\linewidth]{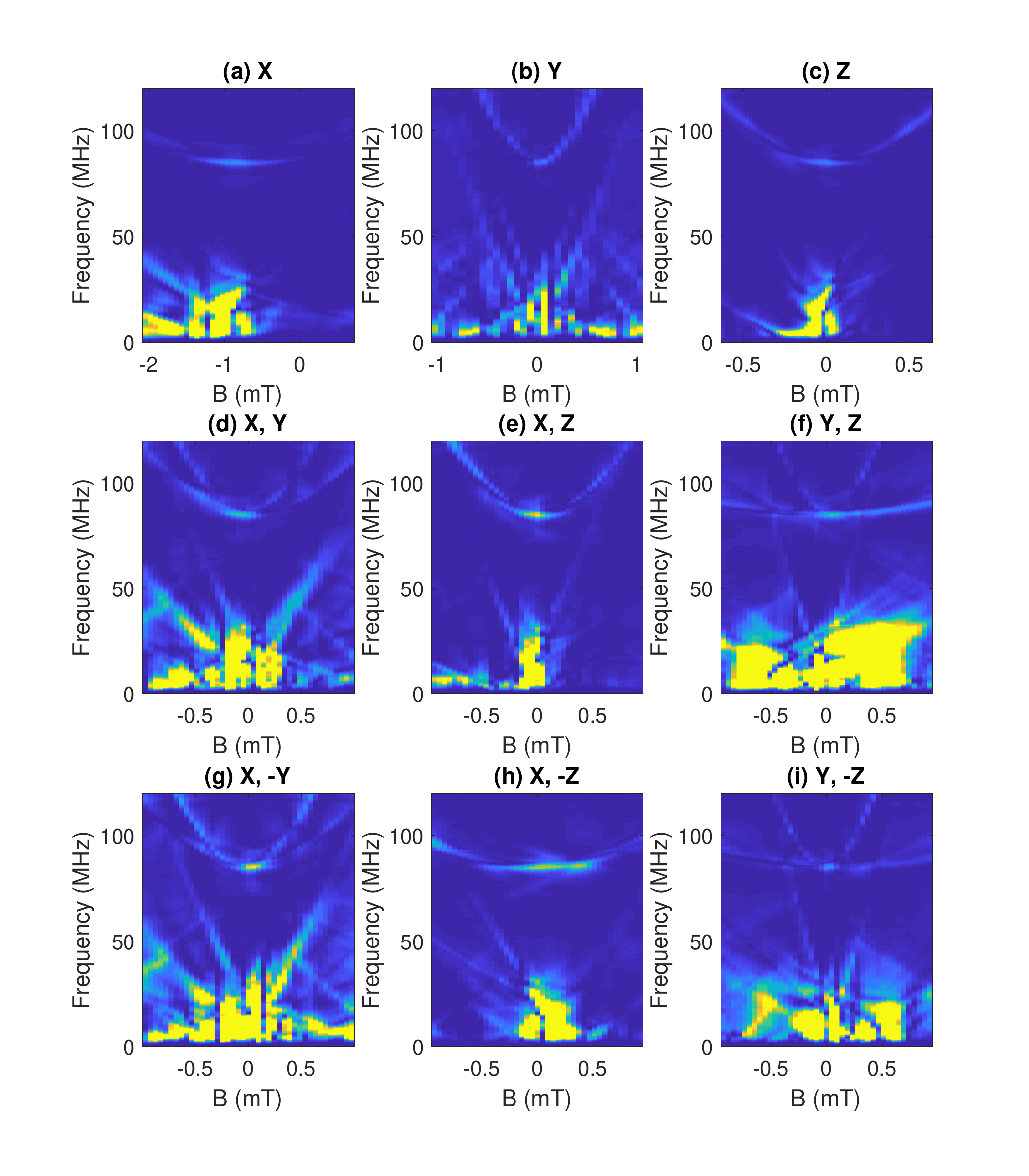}
\caption{\label{fig:85MHz_noline} The same Raman-heterodyne data as for Fig.\ \ref{fig:85MHz} but without the theoretical lines. These plots are included to more clearly show the resonances at $85$ MHz which were obscured by the theory curves in the above figure. Again, field directions are indicated in subfigure titles, with the indicated field directions $X$, $Y$, and $Z$ not coincident with the crystallographic axes. See the text for details.}
\end{figure}

As noted in the main text, Raman-heterodyne spectroscopy was performed in two frequency bands: between 0-120 MHz and 0.6-1.2 GHz. Here the low-frequency Raman-heterodyne data is presented for magnetic field sweeps along nine different directions. The data is shown in two almost identical figures: Fig.\ \ref{fig:85MHz} shows a colormap of transition intensity with the theoretical predictions plotted on top; Fig.\ \ref{fig:85MHz_noline} shows only the raw data, such that transitions which are partially obscured by our model are clearly visible. The theoretical lines correspond to the lowest energy transitions that the crystal-field Hamiltonian predicts for the \term{4}{I}{13/2}$Y_1$ hyperfine manifold. The colormap indicates the linear strength of the Raman-heterodyne signal, with yellow/light blue corresponding to a resonance condition. We note that the large yellow signal at low frequency is likely due to RF pickup, and is saturating the maximum colormap intensity. As a consequence, the transitions at 85 MHz have a somewhat lower intensity. As is evident from the figures, the 85 MHz transitions of both magnetically inequivalent sites agree with the data to within $\sim 2$ MHz for magnetic field sweeps in three dimensions.

During these experiments, some difficulty was experienced in achieving an optimal alignment of the sample inside the RF coil with respect to the magnetic field axes. Therefore, the exact orientation of the sample with respect to the magnet was later deduced by calculating the rotation required for the magnetically inequivalent sites to become degenerate. The transformation to go from the crystallographic $D_1$, $D_2$, and $b$ axes to the $X$, $Y$, and $Z$ coordinate system in the Figs.\ \ref{fig:85MHz} and \ref{fig:85MHz_noline} has the form
\begin{equation}
  R(\theta) = \begin{bmatrix} \cos(\theta) & 0 & \sin(\theta) \\ 0 & 1 & 0 \\ -\sin(\theta) & 0 & \cos(\theta) \end{bmatrix} \begin{bmatrix} 1 & 0 & 0 \\ 0 & 0 & -1 \\ 0 & 1 & 0 \end{bmatrix},
\end{equation}
with $\theta = 174^\circ$. 

\section{Fitting details}

In order to perform the coarse part of the fitting procedure, we employed our own implementation of the basinhopping algorithm \cite{wales_global_1997} in conjunction with the Sbplx local minimization routine \cite{rowan1990,nlopt}. The polishing fit, which included Raman-heterodyne data, instead used simulated annealing followed by a single Sblpx local optimization to obtain the final parameters. For the hyperfine portion of the calculation we used an intermediate coupled effective Hamiltonian \cite{carnall_systematic_1989} in a basis spanning only the \term{4}{I}{15/2} and \term{4}{I}{13/2} states. This lowered the dimension of the hyperfine Hamiltonian from 2912 to 224. Since a separate Hamiltonian had to be diagonalized for each magnetic field orientation, this reduction was essential to making the optimization computationally feasible. 

In all fitting calculations, a number of higher order free-ion parameters were held fixed to the values determined by \cite{carnall_systematic_1989} for Er$^{3+}$:LaCl$_3$; for completeness, these values are provided in Tab.\ \ref{tab:erlacl3_param}.
\begin{table}[tb!]
\caption{\label{tab:erlacl3_param} Parameters for two electron Coulomb ($\alpha$, $\beta$, $\gamma$), three electron Coulomb ($T^2$, $T^3$, $T^4$, $T^6$, $T^7$, $T^8$), as well as higher order spin-dependent ($M^{\text{tot}}$, $P^{\text{tot}}$) interactions, held fixed throughout all calculations at values reported by Ref.\ \cite{carnall_systematic_1989}.}
  \begin{tabular}{cc}
    \hline \hline
    Parameter & Value (cm$^{-1}$) \\
    \hline

    $\alpha$         &   17.79 \\ 
    $\beta$          &  -582.1 \\ 
    $\gamma$         &    1800 \\ 
    $T^2$            &     400 \\ 
    $T^3$            &      43 \\ 
    $T^4$            &      73 \\ 
    $T^6$            &    -271 \\ 
    $T^7$            &     308 \\ 
    $T^8$            &     299 \\ 
    $M^{\text{tot}}$ &    3.86 \\ 
    $P^{\text{tot}}$ &     594 \\ 

    \hline \hline
  \end{tabular} 
\end{table}

To perform the polishing fit, the literature ground and excited state spin Hamiltonians \cite{sun_magnetic_2008,chen2018} were evaluated at a variety of magnetic field directions. These were sampled from the parametric spiral
\begin{equation}
  \mathbf{B} = B_0 \begin{bmatrix} \sqrt{1-t^2} \cos(6\pi t) \\ \sqrt{1-t^2} \sin(6\pi t) \\ t \end{bmatrix}, \quad t \in [-1, 1],
  \label{eqn:b_spiral}
\end{equation}
with $B_0$ the overall magnitude of the field vector. The fit then included the following 95 data points:
\begin{itemize}
  \item 35 electronic levels up to \term{2}{H}{11/2}$F_1$.
  \item 12 data points for the hyperfine splittings of the \term{4}{I}{15/2}$Z_1$ level, sampled at equally spaced intervals according to Eq.\ \eqref{eqn:b_spiral}, with $B_0 = 0.05$ Tesla. 
  \item 12 data points for the Zeeman splittings of the \term{4}{I}{13/2}$Y_1$ level, also according to Eq.\ \eqref{eqn:b_spiral}, with $B_0 = 0.05$ Tesla. 
  \item 4 data points from low-frequency Raman-heterodyne measurements - here we had sufficient experimental data to calculate a three dimensional curvature tensor of transition energy with respect to a magnetic field. The four data points were added to the crystal-field fit at field vectors of $\begin{bmatrix}0 & 0 & 0 \end{bmatrix}$, $\begin{bmatrix} 0.5 & 0.0 & 0.0 \end{bmatrix}$, $\begin{bmatrix} 0.0 & 0.5 & 0.0 \end{bmatrix}$, and $\begin{bmatrix} 0.0 & 0.0 & 0.5 \end{bmatrix}$ (with axes $\begin{bmatrix} D_1 & D_2 & b\end{bmatrix}$ and values in mT).
  \item 32 data points from high-frequency Raman-heterodyne data - this dataset consisted of magnetic field sweeps along the crystallographic $D_2$ axis. Therefore, we fit second order polynomials to the transition energy curvature with respect to magnetic field along the $D_2$ direction. For the crystal-field fit, these polynomials were sampled at field values of $0.0$, $0.3$, and $0.5$ mT. A total of 10 transitions were fit using this method. Two additional transitions had curvatures that were not well approximated by a second order polynomial as they had a very steep gradient around zero field. For these, the crystal-field fit was only performed at a single magnetic field strength of $0.5$ mT. Fig.\ 1 (a) of the main text shows all of these transitions, with a subset of the transitions also visible in the wide-frequency scan shown in Fig.\ 1 (b). 
\end{itemize}

In Tab.\ \ref{tab:e_summary} we summarize the energies of the crystal-field levels up to the \term{2}{H}{11/2}$F_1$ level. These are shown with respect to the experimental energies determined by Doualan \etal \cite{doualan_energy_1995} using site-selective excitation and fluorescence spectroscopy. We see that the maximum deviation is 36 cm$^{-1}$ (\term{4}{I}{15/2}$Z_6$), although Doualan \etal indicated that the assignment of this transition was uncertain. Ignoring levels with uncertain assignments, we obtain a standard deviation of 13.8 cm$^{-1}$ for the differences between our model and the reported experimental values.
\begin{table}[tb!]
   \caption{\label{tab:e_summary} Theoretical energy levels up to the
   \term{2}{H}{11/2} multiplet of Er$^{3+}$:Y$_2$SiO$_5$ site 1 calculated from our
   crystal-field model. For comparison, the experimentally determined values of
   energy levels reported by Doualan \emph{et al}.\ \cite{doualan_energy_1995} have been
   included.  All quoted values are in cm$^{-1}$.  Question marks denote
   uncertain assignments as indicated by the authors of Ref.\
   \cite{doualan_energy_1995}.}
  \begin{tabular}{ccccc}
    \hline \hline
    Multiplet & State & Theory & Experiment & Difference  \\
    \hline
    
    \term{4}{I}{15/2}   & Z$_1$    &     15 &      0 &    -15  \\
                        & Z$_2$    &     47 &     42 &     -5  \\
                        & Z$_3$    &     75 &     86 &     11  \\
                        & Z$_4$    &    130 &    104 &    -26  \\
                        & Z$_5$    &    199 &    172 &    -27  \\
                        & Z$_6$    &    388 &   424? &     36  \\
                        & Z$_7$    &    462 &    481 &     19  \\
                        & Z$_8$    &    508 &    513 &      5  \\
     \term{4}{I}{13/2}  & Y$_1$    &   6522 &   6507 &    -15  \\
                        & Y$_2$    &   6560 &   6547 &    -13  \\
                        & Y$_3$    &   6583 &   6596 &     13  \\
                        & Y$_4$    &   6640 &   6623 &    -17  \\
                        & Y$_5$    &   6777 &   6798 &     21  \\
                        & Y$_6$    &   6833 &   6852 &     19  \\
                        & Y$_7$    &   6867 &   6871 &      4  \\
     \term{4}{I}{11/2}  & A$_1$    &  10206 &  10193 &    -13  \\
                        & A$_2$    &  10236 & 10271? &     35  \\
                        & A$_3$    &  10267 & 10292? &     25  \\
                        & A$_4$    &  10339 & 10308? &    -31  \\
                        & A$_5$    &  10381 &  10369 &    -12  \\
                        & A$_6$    &  10398 & 10383? &    -15  \\
     \term{4}{I}{9/2}   & B$_1$    &  12348 &  12360 &     11  \\
                        & B$_2$    &  12463 & 12460? &     -4  \\
                        & B$_3$    &  12532 & 12528? &     -5  \\
                        & B$_4$    &  12600 &  12612 &     12  \\
                        & B$_5$    &  12666 &  12650 &    -16  \\
     \term{4}{H}{9/2}   & D$_1$    &  15190 &  15169 &    -21  \\
                        & D$_2$    &  15228 &  15220 &     -7  \\
                        & D$_3$    &  15342 & 15360? &     19  \\
                        & D$_4$    &  15384 &  15382 &     -2  \\
                        & D$_5$    &  15492 &  15498 &      6  \\
     \term{4}{S}{3/2}   & E$_1$    &  18264 &  18267 &      3  \\
                        & E$_2$    &  18377 &  18372 &     -5  \\
                        & E$_3$    &  19082 &  19091 &      9  \\
     \term{2}{H}{11/2}  & F$_1$    &  19116 &  19116 &      0  \\
    
    \hline \hline
  \end{tabular} 
\end{table}

Table \ref{tab:g_sh_summary} shows the hyperfine splittings of the ground state at zero field as predicted by our crystal-field model in conjunction with the splittings predicted using the spin Hamiltonian parameters from Ref.\ \cite{chen2018}. We note that the fitting did not directly include the values shown here, since magnetic field vectors were chosen using Eq. \eqref{eqn:b_spiral} which always had $B_0 = 0.05$ Tesla. This avoided any potential uncertainty in the spin Hamiltonian parameters at zero field (which were obtained by electron-paramagnetic resonance and therefore necessarily only from measurements with a non-zero external magnetic field). The standard deviation for the differences between the two models was 51.2 MHz. 

Finally, in Tab.\ \ref{tab:rh_summary} we show the differences between the crystal-field model and the transition energies measured using Raman-heterodyne spectroscopy. For the transitions included here, at the magnetic field values indicated in the table, the standard deviation amounted to 5.3 MHz. 

In order to estimate the uncertainties of our fitted parameters we used the Markov Chain Monte-Carlo technique to sample the posterior probability distribution \cite{aster2011}.  We completed a total of $3 \times 10^6$ trials with $2917230$ accepted steps. From these we discarded the first $5 \times 10^4$ elements to allow the algorithm to ``burn in''. From all subsequent points, we used every $1000$th element to ensure that consecutive samples were not correlated. The step size of individual iterations was tuned to achieve an acceptance rate of $\sim 10\%$, within the range generally recommended for this procedure \cite{aster2011}.

We note that the primary limitation of this method for evaluating parameter uncertainties is that it assumes that we correctly weight the $\chi^2$ contributions in accordance with the experimental uncertainties of each data element. Consequently if a substantially different weighting were selected, say to preferentially fit to electronic levels over hyperfine splittings, the resulting crystal-field parameters would deviate more than the indicated uncertainties. This would accordingly be reflected in the standard deviations given for Tabs.\ \ref{tab:e_summary}, \ref{tab:g_sh_summary}, and \ref{tab:rh_summary}. 

%\bibliography{eryso_crystal_field}
%

\onecolumngrid

\begin{table}[tb!]
  \caption{\label{tab:g_sh_summary} The ground state hyperfine level energies at zero field as predicted by the crystal-field model. These are compared with the transition energies predicted by the spin Hamiltonian from  Ref. \cite{chen2018}}
  \begin{tabular}{cccc}
    \hline \hline
    Level & Crystal-field model (GHz) & Spin Hamiltonian (GHz) & Difference (MHz) \\
    \hline
    $1$ & $0.0000$ & $0.0000$ & $0.0$ \\ 
    $2$ & $0.0000$ & $0.0000$ & $0.0$ \\ 
    $3$ & $0.8736$ & $0.8643$ & $9.3$ \\ 
    $4$ & $0.8737$ & $0.8645$ & $9.2$ \\ 
    $5$ & $1.6942$ & $1.6455$ & $48.7$ \\ 
    $6$ & $1.7037$ & $1.6672$ & $36.5$ \\ 
    $7$ & $2.3355$ & $2.1919$ & $143.7$ \\ 
    $8$ & $2.6710$ & $2.7385$ & $-67.5$ \\ 
    $9$ & $3.1102$ & $3.0662$ & $44.0$ \\ 
    $10$ & $3.5052$ & $3.5237$ & $-18.5$ \\ 
    $11$ & $4.0424$ & $4.0143$ & $28.1$ \\ 
    $12$ & $4.0482$ & $4.0437$ & $4.5$ \\ 
    $13$ & $4.7107$ & $4.7145$ & $-3.8$ \\ 
    $14$ & $4.7107$ & $4.7154$ & $-4.6$ \\ 
    $15$ & $5.3482$ & $5.4199$ & $-71.7$ \\ 
    $16$ & $5.3482$ & $5.4199$ & $-71.7$ \\ 
    \hline \hline
  \end{tabular} 
\end{table}

\begin{table}[tb!]
  \caption{\label{tab:rh_summary} Transition energies measured using Raman-heterodyne spectroscopy compared with the energies predicted from crystal-field theory. This includes both the $85$ MHz data as well as the transitions between $0.6-1.2$ GHz. All but two transition energies are presented for zero external magnetic field; the two exceptions are denoted by a $\dag$ for which an external field of $0.5$ mT was used to calculate the energies. This was done since these transitions exhibit a very steep gradient around zero field, prohibiting an approximation of their curvature using a polynomial model; see text for further details. The transition column indicates which levels were assigned to these energies; level 1 is the ground state, while the lowest hyperfine level of the $^4$I$_{13/2}$$Y_1$ electronic state corresponds to 129.}
  \begin{tabular}{cccc}
    \hline \hline
    Transition & Crystal-field model (MHz) & Raman Heterodyne (MHz) & Difference (MHz) \\
    \hline
    $1-3^{\dag}$ & $873.8$ & $880.0$ & $-6.1$ \\ 
    $2-4^{\dag}$ & $873.4$ & $880.4$ & $-7.0$ \\ 
    $6-8$ & $967.3$ & $953.5$ & $13.9$ \\ 
    $7-9$ & $774.6$ & $775.0$ & $-0.4$ \\ 
    $7-10$ & $1169.6$ & $1169.0$ & $0.6$ \\ 
    $8-10$ & $834.1$ & $824.7$ & $9.4$ \\ 
    $9-11$ & $932.3$ & $931.5$ & $0.7$ \\ 
    $135-137$ & $1011.6$ & $1011.3$ & $0.2$ \\ 
    $135-138$ & $1098.3$ & $1097.7$ & $0.7$ \\ 
    $136-137$ & $666.8$ & $667.7$ & $-1.0$ \\ 
    $136-138$ & $753.5$ & $753.4$ & $0.2$ \\ 
    $137-138$ & $86.7$ & $85.2$ & $1.6$ \\ 
    $142-144$ & $726.5$ & $726.9$ & $-0.4$ \\ 
    \hline \hline
  \end{tabular} 
\end{table}